\documentclass[twocolumn,twoside,slac_two]{revtex4}
\usepackage{graphicx}
\usepackage{fancyhdr}
\begin{document}

\title{On color confinement}


\author{Adriano Di Giacomo}
\affiliation{Pisa University and INFN Sezione di Pisa , ITALY}
\begin{abstract}
The status of the theory of color confinemnt  is discussed.
\end{abstract}

\maketitle

\thispagestyle{fancy}


\section{Introduction}
 Quarks and gluons are the constituents of hadrons and the fundamental fields of the $QCD$ Lagrangean. 
 \begin{equation}
 L_{QCD} = -{1\over 4} Tr[G_{\mu \nu}G^{\mu \nu}] + \Sigma_f  \bar \psi(i D\slash - m_f)\psi
 \end{equation}
 Quarks can be detected by use of electroweak probes at short distances. They have never been observed as free particles. The experimental upper limits  to the existence of free quarks can be found 
 in Particle Data Group reports\cite{PDG}.
 
 The upper limit to the ratio of the quark abundance in nature $n_q$ to that of the nucleons $n_p$
 is ${n_q\over{ n_p}} \le 10^{-27}$ to be compared with the expected value in the Standard Cosmological Model  $10^{-12}$.  
 
 The inclusive cross section $\sigma_p$ to produce a quark or an antiquark  has upper limit  $10^{-40} cm^2$ to be compared with the value expected in the absence of confinement  $\approx 10^{-25} cm^2$.
 
  A suppression factor of $10^{-15}$ can only have as a natural explanation  that  $n_q=0$ and $\sigma_q =0$ due to some symmetry. 
 
 This phenomenon is called color confinement : asymptotic particle in $QCD$ are only colorless.
 
 The obvious question is then: does $QCD$ imply color confinement, and, if so, by what mechanism?
 What is the symmetry which produces confinement?
 
 For the first time it was conjectured in ref\cite{CP} that the Hagedorn limiting temperature \cite{HAG}
 could in fact correspond to a deconfining phase transition from hadrons to a gas of quarks and gluons (the so called Quark Gluon Plasma).
 
 Experiments colliding heavy ions have been set up at CERN and at Brookhaven to detect such transition. It is  not clear what would be the smoking-gun signal for that transition and no clear statement can be done up to now.
 
 Virtual experiments, i.e. numerical simulations of the theory on the lattice have instead demonstrated the existence of a deconfining transition.
  Lattice Montecarlo techniques produce a discretized approximant  to the functional Feynman integral 
  which defines  $QCD$ . If the lattice spacing $a$ is small compared to hadronic scale $\lambda$
  and  $\lambda$ is in turn much smaller than the lattice size $L a$ the numerical estimate can be a good approximation to the  functional integral.
  
  Also $QCD$ at finite temperature can be simulated by similar techniques. The partition function 
  $Tr[e^{-{H\over T}}]$ is equal to the Feynman integral compactified from $0$ to ${1\over T}$
  in the time direction, with periodic boundary conditions in time for bosonic fields, antiperiodic for fermions. In formulae
  \begin{equation}
   Tr[e^{-{H \over T}}] =  \int [d A_{\mu}][d \bar \psi][d \psi] e^{\int d^3x \int_0^{1\over T}L(A,\bar \psi,\psi)}
   \end{equation}
   On lattice 
   \begin{equation}
    T={1\over {aN_t}}
    \end{equation}
      with  $a= a(\beta, m)$ the lattice spacing, $\beta={{2N}\over{g^2}}$ 
   the usual lattice variable.
   The integral in Eq(1) is computed by simulating on a lattice with time extension $L_t$ and space extension $L_s^3$ with   $L_s \gg L_t$ .
   
   Renormalization group arguments give
   \begin{equation}
   a \propto {1\over \Lambda_L} e^{\beta\over {2b_0}}
   \end{equation}
   with  
   \begin{equation}
   b_0 = - {1\over {(4 \pi)^2}}[{11\over 3} N - {2\over 3}N_f]
   \end{equation}
   the (negative) coefficient of the lowest order term of the $\beta$ function.
   
    The negative sign means asymptotic freedom.
    
     $ \Lambda_L$ is the physical scale of the lattice regularized $QCD$.
   
  $T$ is given by Eq(3)
  \begin{equation}
  T =  {\Lambda_L \over L_t} e^{\beta \over {2|b_0|}}
  \end{equation}
  High $T$ corresponds to small $g^2$  (order) , low  $T$  to large  $g^2$ (disorder) , the opposite of what usually happens to ordinary systems in statistical mechanics.
  
  This peculiar fact naturally  brings us  to Duality\cite{KW}\cite{KC}. 
  
  \section{Duality}
     
     Duality is a deep concept in statistical mechanics and field theory. It applies to systems in $(d+1)$ dimensions  which can have topologically non trivial excitations in $d$ dimensions. 
     
     These systems admit two complementary descriptions.
     
     1) A direct description in terms of local fields  $\Phi(x)$, with  $\langle \Phi \rangle$ the order parameters, in which the topological excitations $\mu$ are non local .  This description is convenient in the weak coupling regime $g  <1$
     
     2) A dual description in which  the topological excitations $\mu$ are local fields and $\langle \mu \rangle $ the (dis)order parameters. In this description the original $\Phi$ fields are non local.
     The dual coupling being $g_D \approx {1\over g}$  this description is convenient at large $g$ (strong coupling).
     Duality maps the strong coupling regime of the direct description into the weak coupling of the dual and viceversa.
     
     The prototype system for duality is the $2d$ Ising model in which the field is a dicotomic variable $\sigma = \pm 1$  defined on the sites of a square lattice. The Lagrangean is the sum on nearest  neighbors of a  ferromagnetic coupling  $ S =  - \Sigma_ {ij}  {J\over T} \sigma _i \sigma _j $ . Putting  $g = {T\over J} $ the partition function is  $Z[g ,\sigma] = \Sigma e^S$ .
     
     Identifying one of the coordinate axes ,$x $, with space, the other one $t$ with time  the topological excitations $\mu$ are  kinks and anti-kinks : a kink  $\mu (\bar x,t)$  is a spatial configuration at time $t$ with $\sigma = -1$ at he sites $x < \bar x$ ,  $x= +1$ at  $x \ge \bar x$ , an anti-kink  has opposite signs.
     It is easily shown that also the operator $\mu$ which creates a kink has eigenvalues $\pm 1$.
     
     Duality  for this model is the equality  \cite{KC} 
     \begin{equation} 
     Z[ \beta ,\sigma]  =  Z[\beta ',\mu]
    \end{equation} 
    where $\beta'$ is defined by the equation
    \begin{equation}
    sinh(2\beta ')  =  {1\over {sinh(2\beta)}}
    \end{equation}
    Eq's(7) and(8) summarize what we have said about duality.
    
     Since in $QCD$ the low temperature phase is disordered, it is natural to look  for dual variables
  $\mu$     and for their symmetry , which should be responsible for confinement.
  
  A model example is $N=2$ SUSY $QCD$ \cite{SW} in which dual excitations are known to be monopoles.
  
  Also in  ordinary  $QCD$ monopoles could be the dual excitations \cite{tH75} \cite{M}, and dual superconductivity of the vacuum the mechanism of confinement.
  
  Here the word dual means that the role of electric and magnetic fields are exchanged with respect to ordinary superconductors : monopoles instead of electric charges condense and Meissner effect acts on electric field instead of magnetic.
  
   The pictorial idea is that the chromoelectric  field  acting between a $q \bar q$ pair is channeled into an Abrikosov flux tube by dual Meissner effect  so that the energy is proportional to the distance , which means confinement.  
  
  In this mechanism the deconfining phase transition is an order-disorder transition from a state with $\langle \mu \rangle\neq 0$ to a state with $\langle \mu \rangle = 0$,i.e. from superconducting to normal.
  \section{Lattice $QCD$}
  The only practical known way to study the large distance behavior of $QCD$ , which is related to confinement, is to simulate numerically the theory on a lattice\cite{wil}.
  
  The first question is how to detect confinement and deconfinement on the lattice. The question is non trivial and parallels the same question in experiments. In the absence of quarks (the so called quenched theory) , i.e. in pure gauge theory this question has a clear answer. In the presence of quarks this is not the case any more.
  
  In quenched theory an order parameter can be defined \cite{Pol} which is the vacuum expectation value of the parallel transport along the time direction across the lattice, the Polyakov line $L(\vec x)$,  which is gauge invariant because of periodic boundary conditions.
  \begin{equation}
  L(\vec x) = Tr [P e^{i \int_0^{1\over T} A_0(\vec x,t) dt}]
  \end{equation}
  It can be proved that  $\langle L\rangle = e^{- {\mu_q\over T}}$ with $ \mu_q$ the chemical potential of a quark. In the confined phase $\mu_q$ is infinite and $\langle L \rangle =0$. This can be seen alternatively as follows.
  
  Let $D(x) \equiv \langle L^{\dagger} (\vec x) L(\vec0)\rangle$ be the Polyakov loop correlator. At large distances by cluster property
  \begin{equation}
  D(x) \approx_{x \to \infty} Ce^ {-{ {\sigma x}\over T} } + |\langle L\rangle|^2
  \end{equation}
  On the other hand one has for the potential $V(x)$
  \begin{equation}
  V(x)  = -T lnD(x)
  \end{equation}
  Together with Eq(10) this gives
  \begin{equation}
  V(x) \approx _{x\to \infty} \sigma x  
  \end{equation}
  if  $\langle L\rangle  = 0$
  and
  \begin{equation}
  V(x) \approx _{x \to \infty} const
  \end{equation}
  if $\langle L \rangle \neq 0$.
  
   $\langle \L \rangle$ is an order parameter for confinement, the center of the group $Z_3$ is the corresponding symmetry. 
   
    A standard finite size scaling analysis of its susceptibility $\chi = \int d^3 x D(x)$ allows to establish that the transition is weak first order\cite{B}. The transition temperature it $T\approx 270Mev$.
   
   An alternative order parameter is the vacuum expectation value of an operator $\mu$ which creates a monopole \cite{dg}\cite{ddpp}\cite {5}\cite{6}. If the mechanism of dual superconductivity is at work, $\langle \mu \rangle \neq 0$ means dual superconductivity, $\langle \mu \rangle = 0$ normal vacuum.
   
   If dual superconductivity is the correct mechanism of confinement the behavior of this order parameter should coincide with that of the Polyakov line.
   
    In that case one expects $\langle \mu \rangle \neq 0$ for $T < T_c$, $\langle \mu \rangle=0$ for $T>T_c$.  $\langle \mu \rangle$ is the ratio of two partition functions with the same Boltzman factor $\beta = {{2N}\over {g^2}}$\cite{dp}. 
   
    To detect the phase transition one must go to infinite volume. The dependence on the volume $L_s^3$ of susceptibilities is determined by the critical indexes, which in turn identify the order of the transition and its universality class.
    
   Instead of the order parameter   $\langle \mu\rangle$ itself it proves convenient to use the related susceptibility
   
    $\rho \equiv {\partial\over {\partial \beta}}ln(\mu)$ .  
    
    One has, due to the boundary value $\langle \mu \rangle = 1$  at $\beta= 0$  
   \begin{equation}
   \langle \mu \rangle = e^{\int_0^{\beta} d\beta \rho(\beta}
   \end{equation}
   For $T>T_c$ we find \cite{6}\cite{dp} 
   \begin{equation}
   \rho \approx  -|c| L_s + c' 
   \end{equation}
   which by use of Eq(14) means  in the infinite volume limit  $\langle \mu\rangle \to 0$ (normal vacuum).
   
   For $T < T_c$  $\rho \to$ finite limit as $L_s \to \infty$, which, again by use of Eq(14) implies  $\langle \mu\rangle \neq 0$  (dual superconducting vacuum).
   
   For $T\approx T_c$ we expect scaling , since the correlation length goes large compared to lattice spacing.
   
   Dimensional analysis gives
   \begin{equation}
    \langle \mu\rangle \approx L_s^{\gamma} f({a\over{\lambda}}, {{\lambda}\over L_s})
    \end{equation}
    with $\gamma$ a possible anomalous dimension, and $\lambda$ the correlation length of the order parameter. Approaching $T_c$, when $\tau \equiv (1 - {T\over T_c}) \to 0$,  $\lambda $ diverges as
    \begin{equation}
    \lambda \propto 
    \tau^{-\nu}
    \end{equation}
    so that ${a\over {\lambda}} \approx 0$ can be neglected. The variable $\lambda \over L_s$ can be traded with $\tau L_s^{1\over {\nu}}$ by use of Eq.(17) so that
    \begin{equation}
     \langle \mu\rangle  \approx L_s^{\gamma} g(0,\tau L_s^{1\over {\nu}})
     \end{equation}
     or 
     \begin{equation}
     \rho \approx L_s^{1\over {\nu}} F(\tau L_s^{1\over {\nu}})
     \end{equation}
     A best fit to the data gives for gauge group $SU(3)$  $\nu = {1\over 3}$ which corresponds to first order transition, in agreement with the analysis done with the Polyakov line quoted above.
     This indicates that dual superconductivity can be the mechanism of color confinement.
     
     Including dynamical quarks explicitly breaks $Z_3$ symmetry and the Polyakov line is not an order parameter any more.  
     
     Moreover there is the phenomenon known as string breaking:  instead of increasing the potential energy when pulling apart a static $q-\bar q$ pair the system prefers to create pairs in the form of pions, so that there is no string tension but there can be confinement.
     
     The situation is depicted schematically in Fig.1 for two quark flavors of equal mass $m$. At large values of the mass  quarks decouple and the theory goes to quenched . There the phase transition is first order and well understood. 
     
      At $m = 0$    there is a chiral phase transition where the spontaneously broken chiral symmetry is restored : $\langle \bar \psi \psi \rangle$ is the order parameter. 
      However at non zero values of $m$
     chiral symmetry is explicitely broken.  
     
     The transition line in the figure is defined by the maxima of a number of susceptibilities (  the specific heat  $C_V$, the susceptibility of the Polyakov line, the susceptibility of the chiral order parameter $\langle\bar \psi \psi\rangle$) which all coincide within statistical errors.
     
     The region above the line is conventionally called "deconfined" , the region below it "confined".
     
     The transition can be explored by use of the dual superconductivity order parameter $\langle \mu \rangle$\cite{ddlpp} and the result is that indeed vacuum is a dual superconductor in the region below 
     the transition line , and goes to normal above it. Moreover the transition is consistent with first order
     at least in the region of low masses.
     
     The order of the transition across the critical line in Fig (1) is a fundamental issue in the study of confinement.  As noticed in Sect.1 a natural explanation of observations on the existence of free quarks
     is that the deconfining phase transition is an order-disorder transition : a continuous transition (crossover) would imply continuity, and hence would require an unnatural way of explaining the inhibition factors of order 
     $\approx 10^{-15}$ which are observed in nature.
     \begin{figure}
\includegraphics[width=65mm]{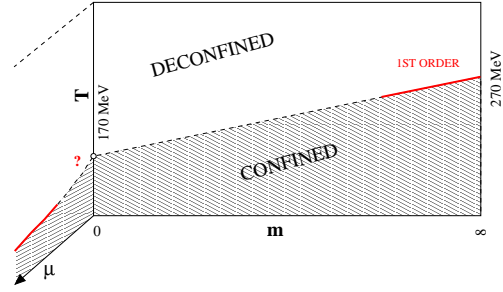}
\caption{The phase diagram of $N_f=2  QCD$. The transition line is defined by the maxima of the specific heat and of the susceptibility of the chiral order parameter. $m$ is the quark mass, $\mu$ the baryon chemical potential.}
\label{Fig.1}
\end{figure}

     An analysis of the chiral transition based on renormalization group and $4-\epsilon$ can be made \cite{PW},
     assuming that the relevant critical degrees of freedom are the scalars and pseudoscalars. 
  
       The result is that the chiral transition is first order for $ N_f  \ge 3$.  
       
       For  $N_f = 2$ two possibilities exist depending
     on the behavior of axial $U_A(1)$ symmetry across the critical point . That symmetry is broken by anomaly at $T=0$ and is expected to be restored at some temperature. 
     
      If $m_{\eta'}$ the mass of the
     singlet pseudoscalar vanishes at $T_c$  then the chiral transition is first order and the same is at $m \neq 0$ in a neighbor of the chiral point $m=0$. 
     
     If instead $m_{\eta'}=0$ at $T_c$ the chiral transition is second order in the universality class of $O(4)$ and a crossover at $m \neq 0$. In this case a tricritical point exists at non zero value of the baryon chemical potential [See. Fig.(1) ]  which could be observed in heavy ion experiments, but has not been observed up to now.
     
     The order of the chiral transition can be investigated on the lattice by looking at the behavior of the specific heat and of other susceptibilities as a function of tha spatial volume.
     
     This analysis requires large amounts of computer time on supercomputers .
      Pioneering work on the subject was inconclusive. More recently a big effort has been put on the problem\cite{ddp} which is still going on. 
      
      The motivation for such an effort is the relevance of this issue to the understanding of confinement mechanisms. All the existing evidence points to a first order transition, or to an order disorder nature of the deconfining transition. This also agrees with the observed behavior of the order parameter $\mu$ for dual superconductivity \cite{ddlpp}.
     More work is needed , however , to fully clarify the situation.
\bigskip % extra skip inserted

\end{document}